\documentclass[conference]{IEEEtran}
\IEEEoverridecommandlockouts
\usepackage{cite}
\usepackage{amsmath,amssymb,amsfonts}
\usepackage{algorithmic}
\usepackage{graphicx}
\usepackage{epstopdf}
\usepackage{textcomp}
\usepackage{xcolor}
\usepackage{mathtools}
\usepackage[ruled,linesnumbered,noend]{algorithm2e}
\usepackage{caption}
\usepackage{tabularray}
\usepackage{subcaption}
\usepackage{booktabs}
\usepackage{enumerate}
\usepackage{rotating}
\usepackage{bbding}
\usepackage{pifont}
\usepackage{wasysym}
\usepackage{xspace}
\usepackage[inline]{enumitem}
\usepackage{bm}

\usepackage[nolist,nohyperlinks]{acronym}
\newcommand{\po}[2]{P^{\mathcal{#1}}_{#2}}

   \usepackage[english]{babel}
    \usepackage{csquotes}

\begin{document}

\title{A Lightweight QoS-Aware Resource Allocation Method for NR-V2X Networks\\
}

\author{\IEEEauthorblockN{Chitranshi Saxena$\textbf{}^1$, Krishna Pal Thakur$\textbf{}^2$, Deb Mukherjee$\textbf{}^1$, Sadananda Behera$\textbf{}^1$, Basabdatta Palit$\textbf{}^1$}
\IEEEauthorblockA{$\textbf{}^1${Department of Electronics and Communication Engineering,  National Institute of Technology, Rourkela} \\
$\textbf{}^2${CEWiT, Chennai}}
}

\maketitle

\begin{abstract}
Vehicle-to-Everything (V2X) communication, which includes Vehicle-to-Infrastructure (V2I), Vehicle-to-Vehicle (V2V), and Vehicle-to-Pedestrian (V2P) networks, is gaining significant attention due to the rise of connected and autonomous vehicles. V2X systems require diverse Quality of Service (QoS) provisions, with V2V communication demanding stricter latency and reliability compared to V2I. The 5G New Radio-V2X (NR-V2X) standard addresses these needs using multi-numerology Orthogonal Frequency Division Multiple Access (OFDMA), which allows for flexible allocation of radio resources. However, V2I and V2V users sharing the same radio resources leads to interference, necessitating efficient power and resource allocation. In this work, we propose a novel resource allocation and sharing algorithm for 5G-based V2X systems. Our approach first groups Resource Blocks (RBs) into Resource Chunks (RCs) and allocates them to V2I users using the Gale-Shapley stable matching algorithm. Power is then allocated to RCs to facilitate efficient resource sharing between V2I and V2V users through a bisection search method. Finally, the Gale-Shapley algorithm is used to pair V2I and V2V users, maintaining low computational complexity while ensuring high performance. Simulation results demonstrate that our proposed Gale-Shapley Resource Allocation with Gale-Shapley Sharing (GSRAGS) achieves competitive performance with lower complexity compared to existing works while effectively meeting the QoS demands of V2X communication systems.

\end{abstract}


\section{Introduction}\label{sec:Intro}
 \ac{V2X} communication, which includes the \ac{V2I}, \ac{V2V}, and \ac{V2P} networks, has garnered significant research attention in recent times due to the growth in the popularity of connected vehicles and autonomous driving~\cite{Garcia2021}. The \ac{V2I} and \ac{V2P} networks are concerned with exchanging road safety, traffic, and infotainment messages with the \acp{RSU} and the pedestrians, respectively. On the other hand, the \ac{V2V} networks are primarily aimed at the inter-vehicle exchange of road safety messages. As a result, the \ac{QoS} requirements, such as latency and reliability,  of \ac{V2V} services are stricter than those of their \ac{V2I} counterparts. For example, the \ac{BLER} requirement of \ac{V2I} users is 0.1 and that of \ac{V2V} users is 0.001~\cite{Korrai2020}. One possible method to address such diverse \ac{QoS} requirements while increasing the data rates is to design efficient radio resource allocation methods, which has been addressed in this work.
 
     To support the diverse \ac{QoS} requirements, the \ac{5G} \ac{NR-V2X} standard (Release 16) uses multi-numerology based \ac{OFDMA}. Thus, while in \ac{4G}, the orthogonal time-frequency \acp{RB} have uniform bandwidths and time duration, those in  5G have different bandwidths and different time duration suitable for the \ac{QoS} requirements of the different services. \ac{NR-V2X} NR-V2X provisions \ac{V2V} communication to occur over the \ac{SL} of existing cellular infrastructure. In other words, the \ac{V2V} communication takes place over the same radio resources assigned to the \ac{V2I} users. The interference inherently associated with the coexistence of \ac{V2I} and \ac{V2V} users in the same radio resource is mitigated using efficient power allocation methods~\cite{Le2017}. 

Efficient radio resource allocation is thus essential to meet the QoS requirements of V2X communications while minimizing interference. In practice, the number of RBs required by a user depends on both the packet size and the Modulation and Coding Scheme \ac{MCS} supported by the user’s link condition. Poor link conditions lower the Signal-to-Interference-Plus-Noise Ratio \ac{SINR}, which in turn requires the use of additional RBs to meet QoS targets. Although 3GPP does not mandate contiguous RB allocation, doing so is advantageous for reducing control channel overhead~\cite{TS38101,TS38214,Dahlamn2016}, especially in Frequency Range 2. This makes the efficient grouping of RBs into Resource Chunks (RCs) \cite{thakur2023qos} an essential task for improving overall performance.

Given the strict QoS requirements and the challenges in efficiently sharing resources between V2I and V2V users, this work proposes a novel resource allocation and resource-sharing algorithm for 5G-based V2X systems.  The proposed algorithm first groups the \ac{OFDMA} \acp{RB} into \acp{RC}~\cite{thakur2023qos} and then selects \acp{RC} for \ac{V2I} users using the stable matching Gale-Shapley algorithm~\cite{gale1962college}. It then allocates power to the \acp{RC} in order to effectuate efficient resource sharing between the \ac{V2I} and the \ac{V2V} users using bisection search~\cite{Le2017RA}. Finally, the Gale-Shapley algorithm is used to pair V2I and V2V users. The quadratic complexity of the Gale-Shapely algorithm makes it considerably lightweight than~\cite{thakur2023qos}, while not compromising on the performance. Simulation results show that our proposed Gale-Shapley Resource Allocation with Gale-Shapley Sharing (GSRAGS) performs at par with the algorithm in~\cite{thakur2023qos}. 

\subsection{Related Works}
 Existing works on \ac{C-V2X} communication mostly focus on two primary aspects (i) resource sharing between \ac{V2I} and \ac{V2V} users, and (ii) power allocation to these users to minimize interference in the shared resources. In addition, some works also discuss the selection of \acp{RB} for \ac{V2I} users. The resource sharing between \ac{V2I} and \ac{V2V} users has been solved using matching algorithms like Hungarian~\cite{Li2018,He2019,Guo2019,GeLi2019,Gyawali2019,Wu2021,Le2017} and Gale-Shapley~\cite{Li2019,Rajan2020}, or using deep neural networks and machine learning~\cite{He2019,Chen2019,Chai2022}. Similarly, power allocation has also been solved either using optimization methods~\cite{Li2019,He2019,Le2017RA,Guo2019,GeLi2019,Wu2021,Rajan2020,Le2017}, heuristics~\cite{Sun2016}  or using deep learning and reinforcement learning techniques~\cite{Chen2019,Gyawali2019,Ron2022}. The selection of resources for the \acp{V2I} is carried out in~\cite{Gyawali2019} using the Hungarian resource allocation method. Authors in~\cite{Aslani2020}, on the other hand, have used MINLP to allocate resources to both \acp{V2I} and \acp{V2V} users. It assumes that orthogonal \acp{RB} are assigned to the \acp{V2I} and the \acp{V2V} users, and there is no resource sharing between them.

In existing works, the primary assumption is that one user requires one \ac{RB} in a scheduling interval, also called the \ac{TTI}. However, one packet may need multiple \acp{RB} in a \ac{TTI} - a method called link adaptation~\cite{Vega2021}. The number of \acp{RB} needed by a user is a function of its packet size and the \ac{MCS} that its underlying link condition can support. \ac{MCS} determines the number of bits that a \ac{RB} can carry. In the face of poor link condition, the user is bound to experience a lower \ac{SINR}, and hence, will be able to support a lower \ac{MCS},  which will increase the number of \acp{RB} needed to serve one packet in a \ac{TTI} while satisfying the given \ac{QoS} constraints. 

The resource allocation algorithm proposed in~\cite{thakur2023qos} addresses link adaptation using a greedy method to allocate multiple \acp{RB} to the \ac{V2I} user to satisfy their data rate and \ac{QoS} requirements. This greedy \ac{RB} allocation, nonetheless, does not assign contiguous \acp{RB},  which can be important for minimizing control channel overhead, especially in higher frequency ranges like Frequency Range 2. So, the authors in~\cite{thakur2023qos} have grouped multiple \acp{RB} into a \ac{RC} and used the maximum bipartite matching Hungarian algorithm to select \acp{RC} for the \acp{CUE}. The authors in~\cite{thakur2023qos} also use the Hungarian algorithm to select the \acp{CUE}-\acp{VUE} pairs which will coexist in the same \acp{RB}.

The Hungarian algorithm has a cubic time complexity \cite{kuhn1955hungarian}, and similarly, the complexity of the algorithm presented in ~\cite{thakur2023qos} scales based on the maximum of the number of \acp{CUE}, \acp{VUE}, and \acp{RC} that can be scheduled in a \ac{TTI}. Specifically, the number of \acp{CUE} and \acp{VUE} that can be scheduled depends on the capacity of the \ac{PUCCH}, which is influenced by factors such as the system's bandwidth, the number of antenna ports, the aggregation level, and the coreset configuration. As a result, the complexity of the algorithm in \cite{thakur2023qos} increases with larger bandwidth or an increase in the number of antenna ports.
 
 With multi-numerology \ac{OFDMA} in \ac{5G}, as the duration of a \ac{TTI} gets smaller, the practical implementations of resource allocation will, therefore, need a more lightweight solution than~\cite{thakur2023qos} for a higher bandwidth system, and also for a system with a higher number of antenna ports. Hence, this work proposes a more efficient approach to resource allocation in V2X systems, leveraging the quadratic complexity of the Gale-Shapley algorithm \cite{gale1962college} for both RC selection and V2I-V2V pairing. 
 

 \subsection{Contributions}
 The major contributions of this work in distinction with existing works are the following:
 \begin{itemize}
     \item We propose a novel resource allocation algorithm for 5G-based V2X communication systems that addresses the diverse QoS requirements of V2I and V2V users while optimizing resource sharing.
     \item The Gale-Shapley stable matching algorithm is leveraged for both the selection of RCs for V2I users and the pairing of V2I and V2V users, providing a lightweight solution with quadratic complexity.
     \item To mitigate interference between V2I and V2V communications sharing the same radio resources, we implement a power allocation mechanism using a bisection search algorithm. 
     \item Simulation results demonstrate that our proposed Gale-Shapley Resource Allocation with Gale-Shapley Sharing (GSRAGS) algorithm achieves performance comparable to existing methods with significantly lower computational complexity
 \end{itemize}

  Rest of the paper is organized as follows. Section~\ref{sec:Sys_Model} discusses the system model while Section~\ref{Sec:Methodology} explains the methodology of how Gale-Shapley algorithm has been used to solve the resource allocation problem. Section~\ref{sec:evaluation} discusses the results while Section~\ref{sec:Conclusion} gives the Conclusion.
\section{System Model}\label{sec:Sys_Model}
\begin{figure}[t]
    \centering
    \hspace{-.4in} \includegraphics[width=0.9\linewidth,trim = {8cm 4cm 9cm 5cm}]{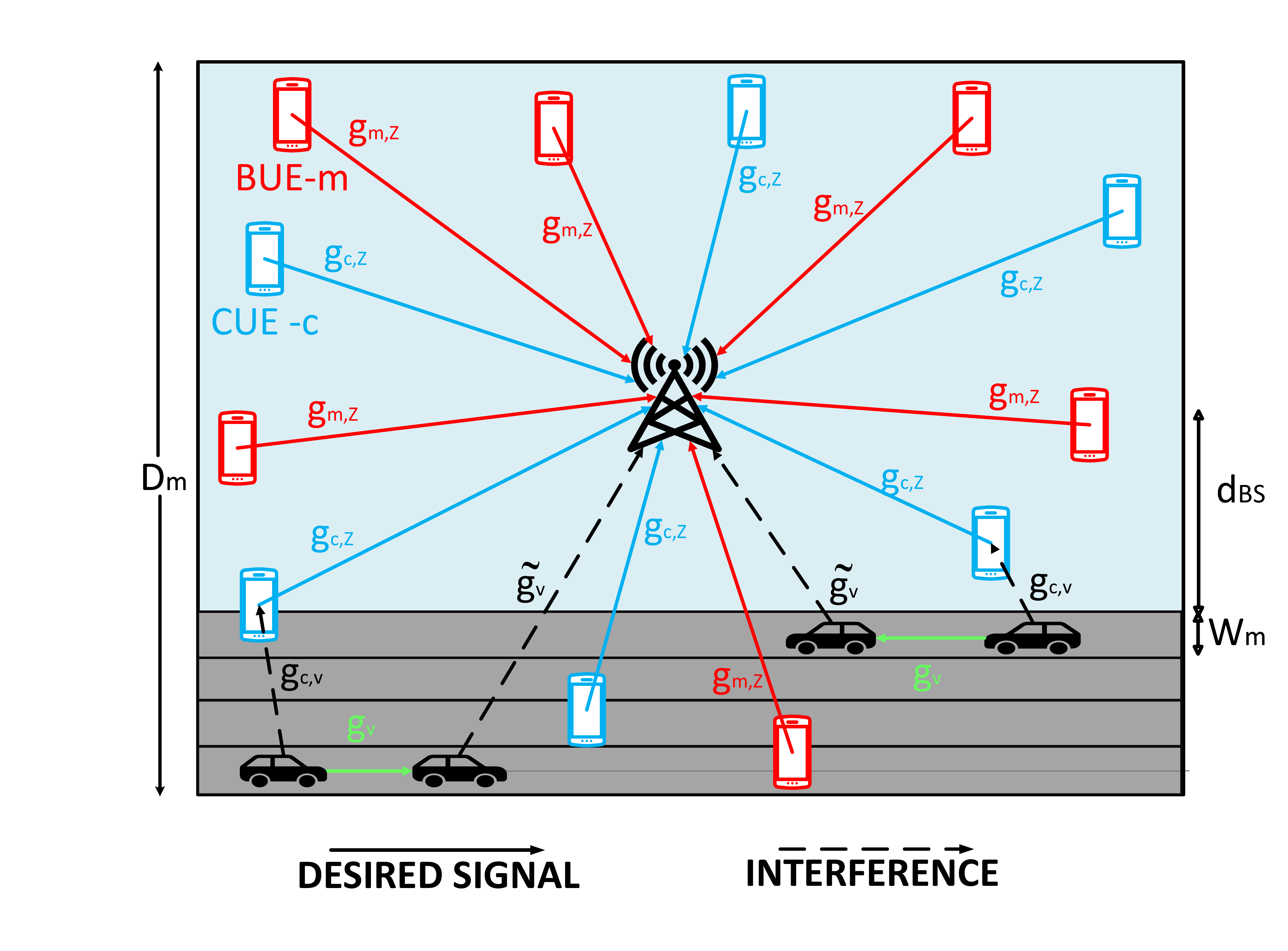}
    \caption{System Model.}
    \label{fig:sys_model}
\end{figure}

\begin{table*}[ht]
\centering
\caption{Symbol Definitions}\label{tab:symbdef}
\begin{tabular}{|p{1cm}|p{6cm}|p{1cm}|p{6cm}| } 
 \hline
 \textbf{Symbol} & \textbf{Definition} & \textbf{Symbol} & \textbf{Definition} \\ 
 \hline 
 $c$ & Index of Cellular User-Equipments (CUEs) & $\gamma^n_i$ & SINR of UE $i$ in the RB $n$, $i\in\{c,v\}$ \\
 \hline
 $v$ & Index of Vehicular User-Equipments (VUEs)& $R_c$ & Rate of CUE $c$ \\
 \hline
 $m$ & Index of Best-effort User-Equipments (BUEs)& $\zeta^n_c$ & Resource Chunk allocation variable\\ \hline
  $\delta_i$& Time-to-live of packets of user type $k$, $k\in\{c,v\}$ &  $x_{c,v}$ & Resource Sharing indicator variable\\ 
 \hline 
  $\beta_i$ & Packet Size of user type $k\in\{c,v\}$   &   $r_0$ & Minimum rate requirement of CUEs \\ \hline
  $\mathcal{B}_i$  & Time-Domain buffer  of user type $i\in\{c,v\}$  &  $D_i$ & Average packet delay of user type $i\in\{c,v\}$\\
 \hline
 $P^\mathcal{C}_i$ & Transmit power of user type $i\in\{c,v\}$ & $p_0$ & Maximum Outage probability of VUEs \\
 \hline
 $P^\mathcal{C}_{max}$ & Maximum transmit power of CUEs & $\gamma_0$ & Minimum SINR Threshold for  VUEs\\ \hline
 $P^\mathcal{V}_{max}$ & Maximum transmit power of {VUEs} &  $C_t$ & Maximum number of users  scheduled  per TTI\\ \hline

\end{tabular}
\end{table*}
In this section, we explain the system model of our work, which considers the underlay mode of resource allocation. An outline of the system model is shown in Fig.\ref{fig:sys_model}. 
\subsection{System Description}
\subsubsection{Scenario and User Traffic} We have considered the uplink of a single \ac{gNB} located at the center of a square service area of $\mathtt{D_m\times D_m}$ sq. km. There are four lanes inside the service area, located to the south of the \ac{gNB} (see Fig. \ref{fig:sys_model}). Inside the service area, the \ac{V2I} users are stationary and include $C$ \acp{CUE} and $M$ \acp{BUE}, and while there are $V$ \ac{V2V} pairs i.e., $2V$ \acp{VUE}. The \acp{V2I} users communicate directly with the \ac{gNB} and are uniformly distributed in the service area, excluding the lanes. The \ac{CUE} traffic is characterized by a delay and a rate constraint and is assumed to be of an infotainment nature. Each \ac{CUE} generates a packet of $\beta_c$ bytes every $\tau_c$ ms, with the number of \acp{CUE} generating a packet in a time slot\footnote{We have considered the time to be divided into fixed-length slots.} following a Poisson distribution with mean $\lambda_c = \frac{\text{Total no. of \ac{V2I} users}}{\tau_c\text{ms}}$. The \ac{V2V} pairs move with a velocity $v$ and are dropped uniformly in the lanes. The \acp{CUE} and \acp{VUE} users have a delay constraint of $\delta_c$ and $\delta_v$ seconds, respectively. In other words, the \ac{CUE} and \ac{VUE} traffic are real-time in nature. The \acp{BUE} traffic, on the other hand, is a non-real-time full buffer traffic, i.e., it always has some data to transmit with no delay constraint.
\begin{table}[h]
\centering
\caption{5G NR Numerologies~\cite{TS38211}}\label{tab:numerology}
\begin{tabular}{|c|c|c|c|}
\hline
\begin{tabular}[c]{@{}c@{}}\textbf{Numerology} \\\textbf{ Index} \\ $\mu$\end{tabular} & \begin{tabular}[c]{@{}c@{}}\textbf{Sub-Carrier} \\ \textbf{Spacing}\\ $\Delta f = 2^\mu\times\mathrm{15 KHz} $\end{tabular} & \begin{tabular}[c]{@{}c@{}}\textbf{Bandwidth} \\ \textbf{of one RB} \\ (\textbf{KHz})\end{tabular} & \begin{tabular}[c]{@{}c@{}}\textbf{TTI}  \\ \textbf{duration} \\ \textbf{(ms)}\end{tabular} \\ \hline
0 & 15 & 180 & 1 \\ \hline
1 & 30 & 360 & 0.5 \\ \hline
2 & 60 & 720 & 0.25 \\ \hline
3 & 120 & 1440 & 0.125 \\ \hline
4 & 240 & 2880 & 0.0625 \\ \hline
\end{tabular}
\end{table}

\subsubsection{Multi-Numerology Resource} To cater to the diverse \ac{QoS} requirements of \acp{CUE}, \acp{VUE}, and \acp{BUE}~\cite{Garcia2021,Korrai2020}, we have used the multi-numerology based \ac{OFDMA} frame structure of \ac{5G}. So, in our system model, the system bandwidth is divided into two orthogonal \acp{BWP}, with each \ac{BWP} being further divided into \acp{RB}. Each \ac{RB} has $14$ \ac{OFDM} symbols, which constitute a \ac{TTI}, in the time domain, and $12$ sub-carriers in the frequency domain. The sub-carrier spacing and the symbol time is determined by the numerology, $\mu$. The first bandwidth part of our system model has $\mu=3$, so its sub-carrier spacing in $2^\mu\times 15=120$ KHz and a \ac{TTI} duration of 0.125 \ac{ms}. The second \ac{BWP} has a numerology $\mu=0$, with the sub-carrier spacing being $15$ KHz and the \ac{TTI} duration of 1 \ac{ms}. As the \acp{CUE} and \acp{VUE} have more stringent latency requirements, hence, they are served in the \ac{BWP}-1, while the \acp{BUE} are served in the \ac{BWP}-2.

\subsubsection{Channel Modeling in Resource Chunks}
In this paper, we have considered the shared or underlay mode of \ac{RB} allocation in which the \acp{CUE} and \acp{VUE} share the \acp{RB}. The channel power gain $g^n_v$ of the \ac{VUE} $v$ in RB-$n$ is given by 
\begin{equation}
  g^n_v = |h^n_{v}|^2\alpha_{v}, \forall v \label{eq:CSI}
\end{equation}

where $h^n_{v}$ is the small-scale fast fading component of the channel and $\alpha_{v}$ is the large-scale fading gain. $h^n_{v}$ is independent and identically distributed (i.i.d) as $\mathcal{CN}(0,1)$. We have considered a block fading channel model such that $h^n_v$ is different in different \acp{RB} in the same \ac{TTI}. Similarly, one can obtain the power gain of the different communication links of Fig. \ref{fig:sys_model} in RB-$n$, i.e., - 
\begin{enumerate*}
    \item $\bm{g^n_{c,Z}}$ - Uplink channel gain between \ac{CUE} $c$ and \ac{gNB} $Z$,
    \item $\bm{g^n_{m,Z}}$ - Uplink channel gain between \ac{BUE} $m$ and \ac{gNB} $Z$,
    \item $\bm{\Tilde{g^n_v}}$ - Interference from \ac{VUE} pair $v$ to the \ac{CUE}-\ac{gNB} communication,
    \item $\bm{\Tilde{g}^n_{c,v}}$ - Interference from the   \ac{CUE} $c$ to the  \ac{VUE} $v$ in \ac{RB}-$n$.
\end{enumerate*}

 Of these, the links with channel power gains, $g^n_{c,Z}$, $g^n_{m,Z}$ and $\Tilde{g^n_v}$ are directly connected to the \ac{gNB}. Therefore, these and the large-scale fading gain, which being a function of user location only varies slowly, can be accurately determined at the \ac{gNB}. However,  the links with channel power gains ${g^n_v}$ and $g^n_{c,v}$ are not directly connected to the \ac{gNB}. So, the corresponding \ac{CSI} reported to the \ac{gNB} has a feedback delay of one TTI, i.e., $T$ milliseconds. The delayed \ac{CSI} feedback and the Doppler shift due to the high vehicular mobility cause the \ac{gNB} to obtain only an estimated channel fading gain $h$ with an error $e$, in these channels. In this work, $h$ is assumed to follow a first-order Gauss-Markov process over a period $T$, \textit{i.e.},
\begin{align}
h=\epsilon \hat{h} + e.
\end{align}
Here $h$ and $\hat{h}$ are the small-scale fading channel gain in the current and previous \acp{TTI}, respectively. The difference, $e$, is considered to be an i.i.d $\mathcal{CN}(0, 1-\epsilon^2)$ variable that is independent of $\hat{h}$. $\epsilon$ is the channel correlation coefficien  between two consecutive \acp{TTI}. It follows Jake's fading model~\cite{Stuber1996}, such that $\epsilon = J_0(2\pi f_d T)$, where  $J_0(.)$ is the zeroth-order Bessel function of the first kind, $f_d = \nu f_c/c$ is the maximum Doppler shift in frequency, $c = 3 \times 10^8$ m/s, $\nu$ is the vehicular speed, and $f_c$ is the carrier frequency.

\subsubsection{Resource chunks and User Rate}
\begin{figure}
    \centering
\includegraphics[width=\linewidth]{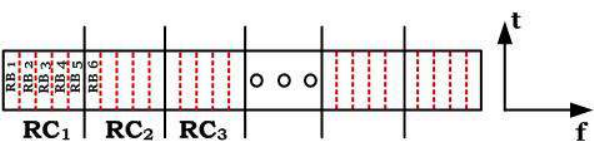}
    \caption{A schematic representation of \acp{RC} for $\eta=5$.}
    \label{fig:enter-label}
\end{figure}
As mentioned in Section \ref{sec:Intro}, in this work, we have considered that multiple \acp{RB} are grouped into a \ac{RC}~\cite{thakur2023qos}, as in shown in  Fig.~\ref{fig:enter-label}. Each \ac{RC} consists of $\eta$ \acp{RB}. Each \ac{CUE} is assigned one \ac{RC} from \ac{BWP}-1.  To control the interference between neighbouring \acp{CUE}, we have assumed that the \acp{gNB} allocates orthogonal \acp{RC}  to the \acp{CUE}. So, we have defined a \ac{RC} allocation variable $\zeta^i_c$ such that $\zeta^i_c =  1$, if \ac{RC} $i$ is allocated to \ac{CUE} $c$, and $\zeta^i_c =  0$, otherwise.

Assuming that the \ac{RC}-$i$ is shared between \ac{CUE} $c$ and \ac{VUE} $v$, the datarate of  $c$  in \ac{RC}-$i$ is given by:
\begin{equation}\label{eq:rateRCcue}
    R_c^i = \eta \times\gamma_c^{n^*},
    \end{equation}
where $\gamma_c^{n^*}$ denotes the minimum \ac{SINR} among all \acp{RB} in \ac{RC}-$i$, and is given by:
\begin{equation}
    \gamma_c^{n^*} = \underset{n\in\{1+\eta i-\eta:1+\eta i\}}{\mathrm{min}} \gamma_c^n.
\end{equation}
Here, $\gamma_c^n$ is the \ac{SINR} of \ac{CUE}-$c$ in \ac{RB}-$n$, such that:
\begin{equation}
  \gamma^n_c =\frac{ \po{C}{c} \alpha_{c,Z}|h^n_{c,Z}|^2}  {\sigma^2+\sum_{v=1}^{V} x_{c,v}\po{V}{v} \Tilde{\alpha}_v|\Tilde{h}^n_v|^2}, 
  \label{eq:SINR_CUE}
 \end{equation}
Similarly we can obtain the datarate of \ac{VUE}-$v$ in \ac{RC}-$i$, with the \ac{SINR} of the \ac{VUE} $v$ in the \ac{RB}-$n$ being given by :
\begin{equation}
\gamma^n_v =\frac{\po{V}{v} \alpha_{v}(\epsilon_v^2 |\hat{h}^n_v|^2 + |e^n_v|^2)}{\sigma^2+\sum_{c=1}^{C} x^n_{c,v} \po{C}{c} (\epsilon_{c,v}^2 |\hat{h}^n_{c,v}|^2 + |e^n_{c,v}|^2) }.
\label{eq:SINRVUE}
\end{equation}
$\po{C}{c}$ and $\po{V}{v}$ denote the maximum transmit powers of $c$ and $v$, respectively, $\sigma^2$  denotes the noise power. 

\subsection{Problem statement}
In this work, our objective is to find (i) the optimal \acp{RC} for \acp{CUE}, and (ii) the optimal \acp{CUE}-\acp{VUE} pairs which maximize the sum rate of the \acp{CUE}, while satisfying their delay and rate constraints, and the delay and coverage probability constraints of the \acp{VUE}. Considering the shared or the underlay mode, we define an indicator variable $z_{c,v}=1$, such that $z_{c,v}=1$ if \ac{CUE} $c$ are and  \ac{VUE} $v$ coexist in the same \ac{RC}, otherwise we set, $z_{c,v}=0$, $\forall c \in \mathcal{C} = \left\{1, 2, \cdots, C\right\}$, and $\forall v \in \mathcal{V} = \left\{1, 2, \cdots, V\right\}$. 
 Therefore, the objective is:
\begin{align}
    \max_{\{\zeta^i_c\},\{z_{c,v}\}, \{\po{V}{v}\},\{\po{C}{c}\}} \sum_i\sum_{c\in\mathcal{C}} R^i_c \zeta^i_c 
 \label{eq:Sumrate}
\end{align}
s.t.
\begin{subequations}
\begin{align}
    &{\sum_{c\in\mathcal{C}}\zeta^i_c \leq 1,\; \forall i, 
    \label{eq:RAconstraint}}\\
     &D_c \leq \delta_c, \; \forall c \,\in\, \mathcal{C};\;
    D_v \leq \delta_v, \; \forall v \,\in\, \mathcal{V} \label{eq:veh_delay_const}\\
    &\sum_i R^i_c \zeta^i_c \geq r_0,\; \forall c \,\epsilon\, \mathcal{C}\label{eq:rate_const}\\ 
    &\mathtt{Pr}\{ \sum_n\gamma^n_v\zeta^c_n x_{c,v} \leq \gamma_0\} \leq p_0,\; \forall v \in \mathcal{V}\label{eq:out_p_const}\\
     &0\leq \po{C}{c} \leq P^\mathcal{C}_\mathrm{max}, \; \forall c \in \mathcal{C}\label{eq:cell_power_const}\\
    &0\leq \po{V}{v} \leq P^\mathcal{V}_\mathrm{max}, \; \forall v \in \mathcal{V}\label{eq:veh_power_const}\\
    &\sum_{c\epsilon \mathcal{C}} z_{c,v} \leq 1,\; z_{c,v} \, \in \, \{0,1\},\; \forall v \in \mathcal{V}\label{eq:resource_sharing_const_1}\\
    &\sum_{v\epsilon V} z_{c,v} \leq 1,\; z_{c,v} \, \in \, \{0,1\},\; \forall c \,\in\, \mathcal{C}\label{eq:resource_sharing_const_2}.
\end{align}
\end{subequations}

The description of the constraints are as follows:
\begin{itemize}
    \item  \textit{\textbf{\ac{RB} allocation constraint}} - (\ref{eq:RAconstraint}) states that only one \ac{RB} can be assigned to only one \ac{CUE}. 
    \item \textbf{\textit{Delay constraint}} - (\ref{eq:veh_delay_const})  ensures that the average packet delays $D_c$ and $D_v$f or a CUE and a VUE must be less than their respective time-to-live values, denoted as $\delta_c$ and $\delta_v$, respectively. 
    \item \textbf{\textit{Minimum Rate constraint}} - (\ref{eq:rate_const}) 
    guarantees that the RB allocation ensures the minimum required rate
    $r_0$ for each CUE, even when resources are shared between CUEs and VUEs.
    \item \textbf{\textit{Coverage Probability constraint}} - (\ref{eq:out_p_const}) ensures that the resource sharing between CUEs and VUEs must meet a reliability condition for VUE communication.  Specifically, $\gamma_0$ required for VUE communication must be met with an outage probability  $p_0$.
    \item \textbf{\textit{Power Allocation Constraint}} - (\ref{eq:cell_power_const}) 
 and (\ref{eq:veh_power_const}) specify that the transmit powers for CUEs and VUEs are upper bounded by $P^\mathcal{C}_\mathrm{max}$ and $P^\mathcal{V}_\mathrm{max}$,  respectively.
    \item \textbf{\textit{Resource Sharing Constraint}} - (\ref{eq:resource_sharing_const_1}) and (\ref{eq:resource_sharing_const_2})  ensure that the RBs of only one CUE are shared with only one VUE.
\end{itemize}

\section{Proposed Algorithm}\label{Sec:Methodology}
In this section, we present the proposed -Gale-Shapley Resource Allocation with Gale-Shapley sharing (GSRAGS) - algorithm. We outline the description of the algorithm in Section \ref{sec:GSRAGS} and its complexity in Section \ref{sec:complexity}.
\subsection{\textbf{Gale Shapley RA with Gale Shapley Sharing}}\label{sec:GSRAGS}
The algorithm executes at the beginning of each \ac{TTI} and has three major steps. (i) First, it allocates resource chunks to the \acp{CUE}. (ii) Secondly, it undertakes power allocation, i.e., for each \ac{CUE}-$c$ which is assigned a \ac{RC} in step (i),  the algorithm allocates power to $c$ and all the \acp{VUE} in the  \ac{RC} assigned to $c$. (iii) Finally, the algorithm pairs the \ac{CUE} and the \ac{VUE} users.
\subsubsection{\textbf{CUE-RC allocation}}\label{sec:GSRAGSstepi}
\begin{figure}[t]
    \centering
    \includegraphics[scale=.5]{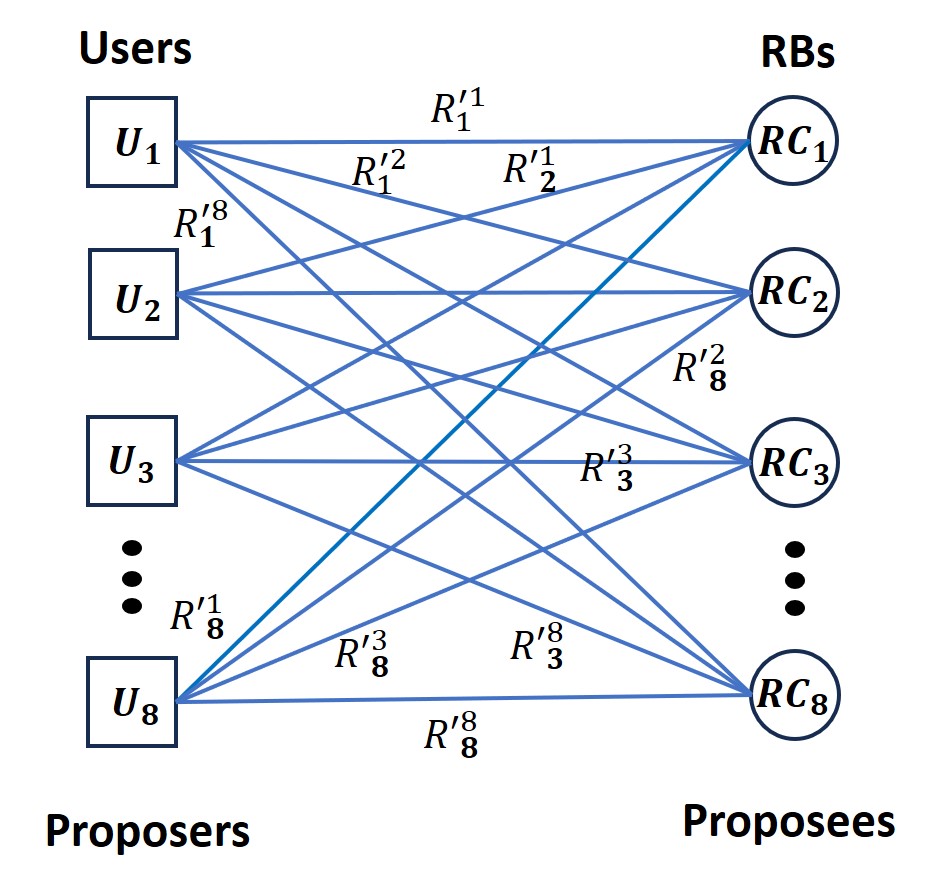}
    \caption{The weighted bipartite graph for RC allocation of \acp{CUE}. The weights of the edges are $R_c^{'i}$, which represents the rate of \ac{CUE}-$c$ in \ac{RC}-$i$ without any interference from the \acp{VUE}.}
    \label{fig:CUERSalloc}
\end{figure}

For assigning \acp{RC} to the \acp{CUE}, we have considered a weighted bipartite graph with two sets of nodes as shown in Fig. \ref{fig:CUERSalloc}. One set of nodes correspond to the total number of \acp{RC}, the other set corresponds to $C_t$, i.e., the maximum number of users that can be scheduled in a \ac{TTI}. As mentioned earlier, $C_t$ depends on the \ac{PUCCH} capacity. The weight of the edge connecting the node-$c$ from one set of nodes to the node-$i$ of the other set of nodes is $R_c^{'i}$, i.e., the rate supported by \ac{CUE}-$c$ in \ac{RC}-$i$, without considering the interference from any \ac{VUE}. 

We have mapped this resource assignment problem, represented by the bipartite graph in Fig. \ref{fig:CUERSalloc}, into a stable matching problem where each set of nodes ranks the nodes in the other set according to some preferences. In our work, each \ac{CUE} ranks the \acp{RC} in descending order of the rates that the \acp{RC} can provide to it. The \acp{RC}, on the other hand, also rank the \acp{CUE} in a descending order of the rates with which it can serve the \acp{CUE}. We have solved this problem using the Gale Shapley algorithm~\cite{gale1962college}, which was famously used to solve college admission and stable marriage problems with a quadratic complexity. In the Gale Shapley algorithm, there are two sets of nodes, one a set of proposers and the other is a set of proposees. The algorithm begins by considering the preference of the proposers and returns a stable matching. In other words, upon applying the Gale-Shapley algorithm, it may be said that no other matching will be more stable than the current one returned. The algorithm is also optimal for the proposers. In this work, we have considered the \acp{CUE} to be the proposers, and, therefore, the \ac{CUE}-\ac{RC} matching is stable and also optimal for the \acp{CUE}. The output of this step is $\zeta_c^i$, i.e., the \ac{RC} allocation variable for a \ac{CUE}-\ac{RC} pair.
\begin{algorithm}[ht]
\caption{Proposed \textbf{GSRAGS} algorithm} \label{alg:HRAHS}

\textbf{INPUT:} Buffers  $\mathcal{B_C}$ and $\mathcal{B_V}$ in which the \ac{CUE} and \ac{VUE} packets are stored in order of their time-to-live, respectively.\\
\textbf{OUTPUT:} (i) a stable \ac{RC}-\ac{CUE} allocation $\zeta^{j*}_c$ and (ii) a stable optimal pairing $x^{c,v}$ between \acp{CUE} and \acp{VUE}\; 
\For{$t\in\{1,2,\cdots T\}$}{
\If{$C_t>\mathtt{len}(\mathcal{B_C})$\label{algo2Line:shortBuffer}}{Append $\{C_t-\mathtt{len}(\mathcal{B_C})\}$ null users to $\mathcal{B_C}$\;}
\Else{\If{$C_t\leq\mathtt{len}(\mathcal{B_C})$}{Use first $C_t$ users from $\mathcal{B_C}$ \;}}
\For {$c \in \mathcal{B_C}[1:C_t]$\label{algo2Line:takeCt}}{
 Compute  rate $
  R^{'i}_c$ of \ac{CUE}-$c$ in \ac{RC}-$i$ without considering interference from the \ac{VUE} in the same way as  (\ref{eq:rateRCcue})\label{algo2line:calculateCSI}\;}
\textbf{STEP 1:} Apply the Gale-Shapley algorithm to determine the stable matching between $C_t$ CUEs in $\mathcal{B_C}$ and the $C_t$ RCs, aiming to maximize the sum rate of the \acp{CUE}, i.e., $\sum_i\sum_c R_c^{'i}$, while ensuring a \ac{BLER} of 0.1.\\
Add scheduled \acp{CUE} to the list $\mathcal{S_C}$\;
 \For{$c\in S_c$}{
\For{$v \in \mathcal{V} = \{1,2,\cdots,V\} $}{
For the pair  $\{c,v\}$ obtain the optimal power allocation ${P^*_v, P^*_c}$ from (\ref{eq:opt_power_alloc}), which would maximize the \ac{CUE} rate $R_c^i$, where $R_c^i$ is calculated considering interference as in (\ref{eq:rateRCcue}).\label{algo2line:poweralloc}\;
Fix the $R^i_{c}$ corresponding to ${P^*_c, P^*_v}$\label{algo2line:recalRc}\;
\uIf{$R^i_{c}<r_0$} {$R^i_{c} = -\inf$\;}
}}
\textbf{STEP 2:}Use Gale Shapley algorithm to find optimal pairing $x_{c,v}$ between \acp{CUE} and \acp{VUE} based on $R^i_{c}$\label{algo2Line:Hungarianforpairing}\;
\Return
}
 \end{algorithm}
\subsubsection{\textbf{Optimal Power Allocation}}
 Once the \acp{RC} are chosen for the \acp{CUE}, the algorithm then allocates power to each \ac{CUE}-\ac{VUE} pair in the \ac{RC} assigned to the corresponding \ac{CUE} using $\zeta_c^i$. Here, it considers only those \acp{CUE} which are scheduled in the step (i).
 The optimal power allocation problem can be defined as \cite{Le2017}: \begin{align}
&(P^*_c,P^*_v)= \underset{\po{C}{c},\po{V}{v}}    {\mathrm{argmax}}\ R_{c}^i\label{eq:opt_power_alloc}\\
\text{s. t. \ } &(\ref{eq:out_p_const}),   \text{(\ref{eq:cell_power_const})\ and (\ref{eq:veh_power_const})) are satisfied.}\nonumber
\end{align}
Here, $R_c^i$ is given by (\ref{eq:rateRCcue}). This optimization algorithm is carried out over all \acp{VUE} for each \ac{CUE}-$c$. In other words, for $\zeta_c^i=1$, the objective of this step is to choose the power for \ac{CUE}-$c$ and \ac{VUE}-$v$ in such a way that the minimum rate among all the \acp{RB} of \ac{RC}-$i$ is maximized, while satisfying the constraints of (\ref{eq:out_p_const}),   (\ref{eq:cell_power_const}) and (\ref{eq:veh_power_const}). 
\subsubsection{\textbf{CUE-VUE Pairing}}
After the power allocation, we now have a new weighted bipartite graph, where the two sets of nodes are the scheduled \acp{CUE} and the \acp{VUE}. As before, if the numbers mismatch then we have extended the corresponding set with dummy users. This bipartite matching algorithm is also solved with the Gale Shapley algorithm to achieve a lower complexity. The proposers being the \acp{CUE}, the algorithm is optimal for the \acp{CUE}. The output of this step gives the pairing variable $x_{c,v}$ which is set 1 if \ac{CUE}-$c$ is paired with \ac{VUE}-$v$. 
\subsubsection{Scheduling of the \acp{BUE}}
For GSRAGS, we have assigned the \acp{BUE} to \ac{BWP}-2 with a numerology of $\mu=0$, i.e., in this \ac{BWP}, the \ac{RB} have a bandwidth of 180 KHz and a TTI duration of  1ms. The \acp{BUE} transmit full-buffer traffic, i.e., they always have some data to send. The \acp{BUE} are scheduled according to the Max C$\backslash$I algorithm \cite{dahlman2013}, i.e., the \ac{BUE}-$m$ is assigned to that \ac{RB} in which it gets the highest rate.

 \subsection{Computational Complexity Analysis }\label{sec:complexity}
 The Hungarian algorithm employed in the previous works~\cite{Li2018,He2019,Guo2019,GeLi2019,Gyawali2019,Wu2021,Le2017,thakur2023qos} have a cubic complexity. For example, the  \textit{\ac{HRAHS}} scheme of~\cite{thakur2023qos}, which uses Hungarian algorithm for \acp{CUE} resource allocation and Hungarian algorithm for \ac{CUE}-\ac{VUE} matching has a complexity of $\mathcal{O}(\mathtt{max}(C_t,N_V, N_{RC})^3)$, where $C_t$ and $N_V$ are the numbers of \acp{CUE} and \acp{VUE} that are scheduled in a \ac{TTI}, respectively, and $N_{RC}$ denotes the number of \acp{RC} available.  

 In contrast, the proposed GSRAGS scheme uses the Gale-Shapley method for resource allocation to CUEs, which has a lower complexity of $O(n^2)$ \cite{gale1962college}. In GSRAGS, the Gale-Shapley algorithm is first applied to match \acp{CUE} to \acp{RC}. However, to execute the Gale Shapley algorithm the number of \acp{RC} should be same as the number of \acp{CUE}. In a high bandwidth system, if the number of \acp{RC} is much more than $C_t$, i.e., the maximum number of users that can be scheduled in a \ac{TTI} then $\{N_RC-C_t\}$ number of dummy users should be added to the user side. Similarly, if the number of \acp{RC} is less than $C_t$, then $\{C_t-N_{RC}\}$ number of dummy \acp{RC}   should be added to the \ac{RC} set. Thus, when Gale Shapley algorithm is applied, the complexity of the \ac{CUE}-\ac{RC} assignment becomes $\mathcal{O}(\mathtt{max}(C_t, N_{RC})^2)$. Similarly, the complexity of  \ac{CUE}-\ac{VUE} pairing with Gale Shapley algorithm is $\mathcal{O}(\mathtt{max}(C_t, N_{V})^2)$. Therefore, the overall complexity of GSRAGS is $\mathcal{O}(\mathtt{max}(C_t,N_V, N_{RC})^2)$. 
\section{Results and Discussion}\label{sec:evaluation}
In this section, we begin by explaining the simulation parameters, followed by a discussion of the obtained results.
\subsection{Simulation Parameters}
We have considered a $\mathrm{1\ km \times 1\ km}$ square area with the \ac{gNB} positioned at the center. Four lanes, each of $4m$ is located to the south of the \ac{gNB} inside its service area. The \acp{CUE} and \acp{BUE} are distributed according to a uniform distribution in the simulation area, excluding the lanes. The \acp{CUE} and \acp{BUE} represent stationary \ac{V2I} users. Ten \acp{VUE} pairs, all moving at 50 kmph, are uniformly placed on the lanes. The simulation parameters are taken from 3GPP TR 37.885 \cite{TS38211} and are tabulated in presented in Table \ref{tab:sim_param}. We have considered a total system bandwidth of $50$ MHz. As mentioned earlier, we have divided the bandwidth into two  parts, where \ac{BWP}-$1$ uses numerology $\mu=3$, and \ac{BWP}-$2$ uses numerology $\mu=0$. So, each \ac{RB} in  \ac{BWP}-$1$ has a bandwidth of 1440 KHz. We have further assumed that each \ac{RC} is made of four \acp{RB}. The carrier frequency in \ac{BWP}-$1$ is 28 GHz and that in \ac{BWP}-$2$ is 2GHZ. We have simulated this scenario for $5000$ time slots, and the results are averaged over $10$ different runs.  To establish the performance of our proposed GSRAGS algorithm, we have considered two baselines - (i) the \ac{HRAHS} algorithm of \cite{thakur2023qos}, and (ii) another algorithm which uses Gale Shapley matching for \ac{CUE} resource allocation and Hungarian matching for \ac{CUE}-\ac{VUE} pairing. We have compared these algorithms  based on key metrics such as packet loss ratio, delay, sum-rate, the number of RBs required, and the \ac{QoS}-constrained capacity. For a given \ac{VUE} density, we have defined the \ac{QoS}-constrained capacity as the number of \acp{CUE}, such that 95\% of these users have their delay constraint and packet loss constraint (less than 2\%) satisfied.

\subsection{Results}
\begin{table}[t]
\centering
\caption{Simulation Parameters}\label{tab:sim_param}
\begin{tabular}{|p{5cm} | p{3cm} |}
\hline
\textbf{Parameter} & \textbf{Value}  \\
\hline  
Carrier Frequency & 28 GHz  \\
\hline
\ac{gNB} / Vehicle  antenna gain & 8 dBi / 3 dBi \\
\hline
Simulation Area / Lane width  & 1 km$\times$1 km / 4 m \\
\hline
 Number of Lanes & 4 lanes in each direction \\
\hline
BS / Vehicle Noise Figure &  5 dB /  9 dB  \\
\hline
Noise Power $ \sigma^2$  &  -114 dBm \\
\hline
Vehicle speed &  50 kmph \\
\hline
Minimum spectral efficiency of \acp{CUE} $r_0$ &  0.5 bps/Hz \\
\hline
SINR Threshold for coverage of \acp{VUE} $\gamma_0$ &  5 dB \\
\hline
Outage probability  $p_0$ &  $10^{-3}$ \\
\hline
Maximum \acp{CUE}/\acp{VUE} transmit power $P_c$,$P_v$  &  23 dBm \\
\hline
Shadowing type &  Log-Normal \\
\hline
Shadowing standard deviation V2V / V2I & 4 dB / 7.8 dB\\
\hline
Number of \ac{VUE} Pairs / \acp{BUE} & 10 / 10 \\
\hline
 Path Loss  \acp{CUE} &  $ 32.4 + 20 * log10(f_c) + 30 * log10(d)$ \\ \hline
 Path Loss (V2V link) & $36.85 + 30 * log10(d) + 18.9 * log10(f_c)  $ \\
 \hline
Delay constraint of \acp{VUE} ($\delta_v$)/CUE ($\delta_c$) & 10 msec / 50 msec \\
\hline
Packet size CUE ($\beta_c$) / \acp{VUE}  ($\beta_v$) & 50 bytes / 10 bytes \\
\hline
Packet generation rate of CUEs $\lambda_c$  & Number of \acp{CUE}/20 \\
\hline
Feedback Period $T$ & 0.125 ms\\
\hline
\end{tabular}
\end{table}
\begin{figure}
\centering
     \includegraphics[trim={0cm 0cm 0cm .7cm},clip,width=0.4\textwidth]{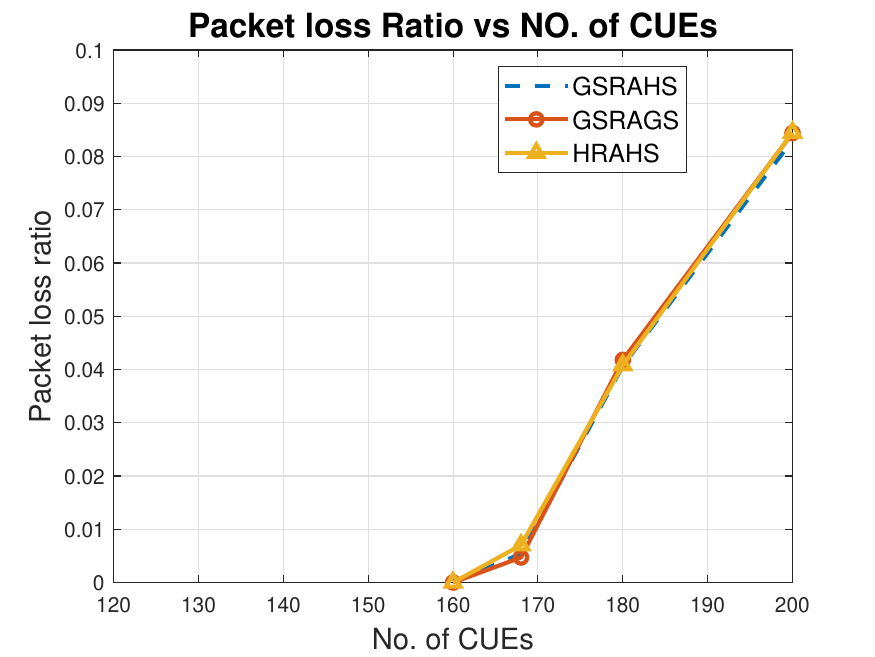}
\caption{ Packet Loss Rate of \acp{CUE} }
\label{fig:CUE_PLR}
\end{figure}
\begin{figure}
       \centering
         \includegraphics[trim={0cm 0cm 0cm .7cm},clip, width=0.4\textwidth]{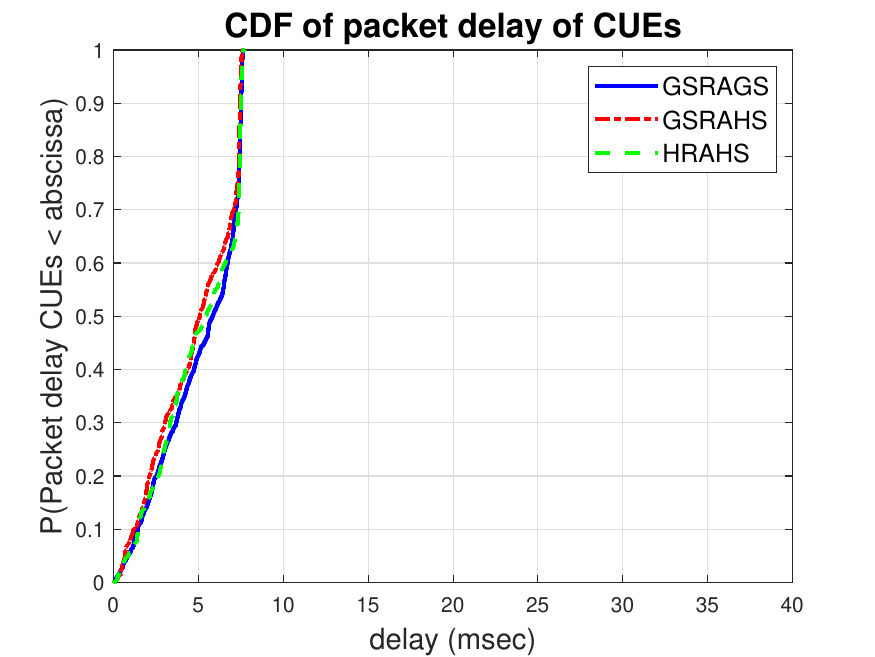}
\caption{Average Delay of \acp{CUE}}
\label{fig:CUE_delay}
\end{figure}

Figure \ref{fig:CUE_PLR} shows how the packet loss ratio of the \acp{CUE} varies as a function of the number of \acp{CUE}. A \ac{QoS}-constrained user is satisfied if the average packet delay it encounters is less than the expiry deadlines and if the packet loss rate is less than two percent. It is observed from Fig. \ref{fig:CUE_PLR} that  the baseline algorithms HRAHS, GSRAHS, and the proposed GSRAGS,  support a \ac{PLR} of two percent for 172 users in the system. Figure \ref{fig:CUE_delay} shows the average packet delay performance of the \acp{CUE} for 172 users. It is observed that although the delay limit of the \acp{CUE} is 50 ms, all three algorithms provide an average delay of less than eight milliseconds.
\begin{figure}[h]
      \centering
         \includegraphics[trim={0cm 0cm 0cm .75cm},clip,width=0.4\textwidth]{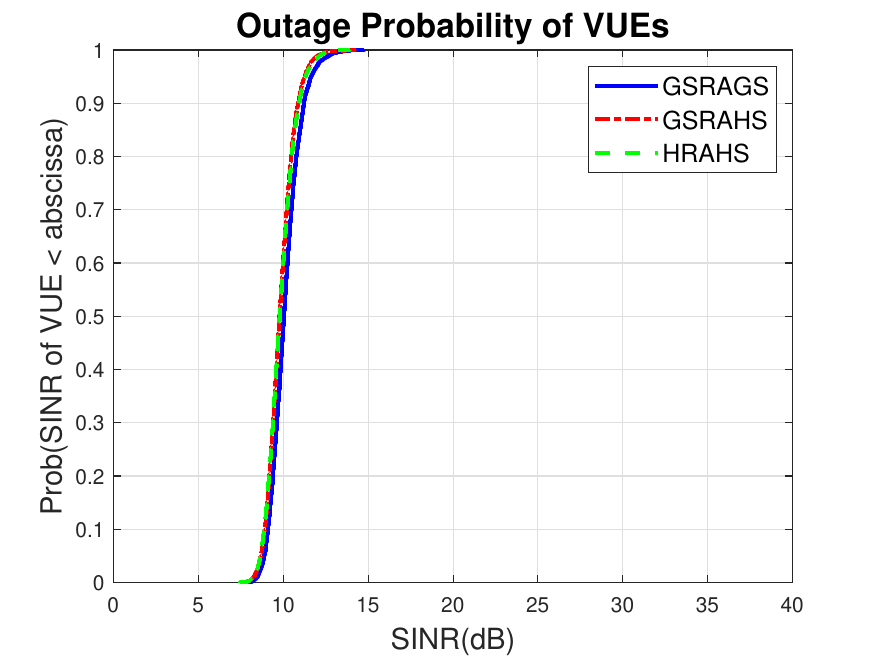}
\caption{ Outage Probability of \acp{VUE}. }
\label{fig:VUE_OP}
\end{figure}
\begin{figure}
   \centering
         \includegraphics[trim={0cm 0cm 0cm .75cm},clip,width=0.4\textwidth]{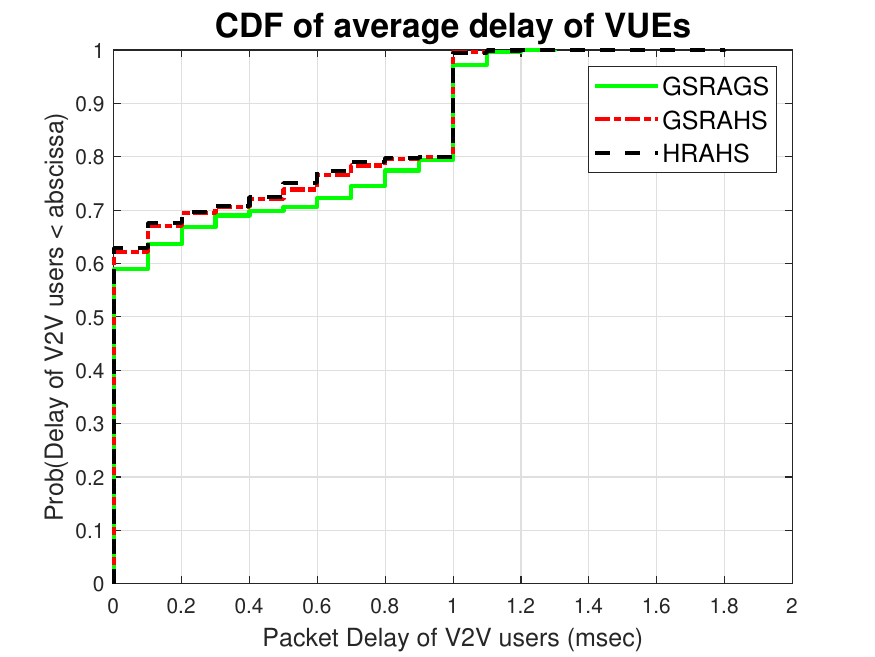}
\caption{CDF of Packet Delay of \acp{VUE}}
\label{fig:VUE_delay}
\end{figure}
  
\begin{figure}[h]
\centering
\includegraphics[width=0.4\textwidth, trim={0cm 0cm 0cm .7cm}, clip]{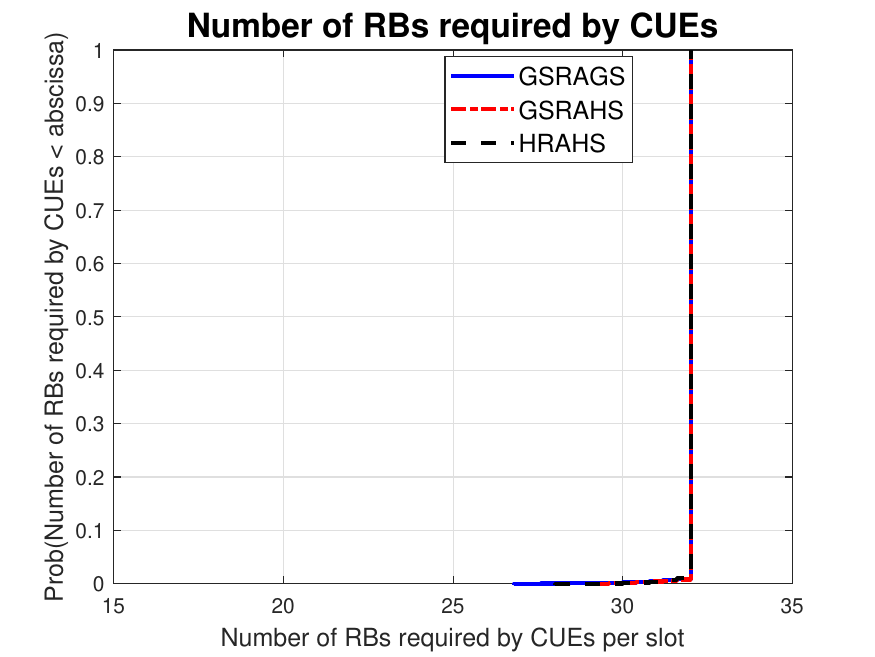}
\caption{Average number of RBs required by CUEs.}
\label{fig:CUE_RB}
\end{figure}

\begin{figure}[h]
\centering
\includegraphics[width=0.4\textwidth, trim={0cm 0cm 0cm .7cm}, clip]{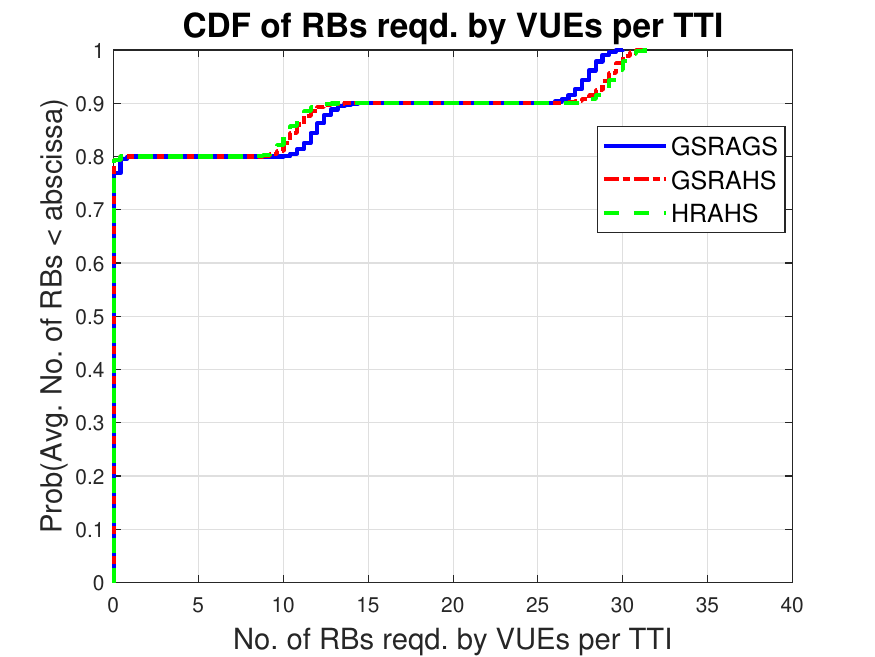}
\caption{Average number of RBs required by VUEs.}
\label{fig:VUE_RB}
\end{figure}

Figures \ref{fig:VUE_OP} and \ref{fig:VUE_delay} shows the CDF of the outage probability of the \acp{VUE} and their average packet delay performance. The \ac{SNR} threshold used for the \acp{VUE} in this work is 5dB and the delay limit is 10 ms. It is observed from Figs. \ref{fig:VUE_OP} and \ref{fig:VUE_delay} that all ten \ac{VUE} users have their output probability as well as their delay deadlines satisfied by the baseline algorithms as well as the proposed algorithm. Therefore, it may be inferred that the proposed GSRAGS renders competitive \ac{QoS} performance in an \ac{NR-V2X} scenario with respect to the existing algorithms, but it will emerge as a better choice because of its lower computational complexity. This claim is further consolidated by Figs. \ref{fig:CUE_RB} and \ref{fig:VUE_RB} which show that the proposed GSRAGS achieve the same performance at a lower complexity but without any increase in the number of occupied \acp{RB}. In this work, we have assumed that a maximum of eight \acp{CUE} can be scheduled in a \ac{TTI}. Accordingly, we have taken the number of available \acp{RC} to be eight. Fig. \ref{fig:CUE_RB} shows that all the three algorithms occupy 32 \acp{RB}, i.e., eight \acp{RC}. Moreover, Fig. \ref{fig:VUE_RB} shows that the number of \acp{RB} required by the \acp{VUE} is also close by. Thus, GSRAGS may be recommended as a light weight resource allocation algorithm which caters to the \ac{QoS} requirements of the real-time traffic carried by the \acp{CUE} and \acp{VUE} in an \ac{NR-V2X} scenario.


\begin{figure}[h]
\centering
\includegraphics[width=0.4\textwidth, trim={0cm 0cm 0cm .64cm}, clip]{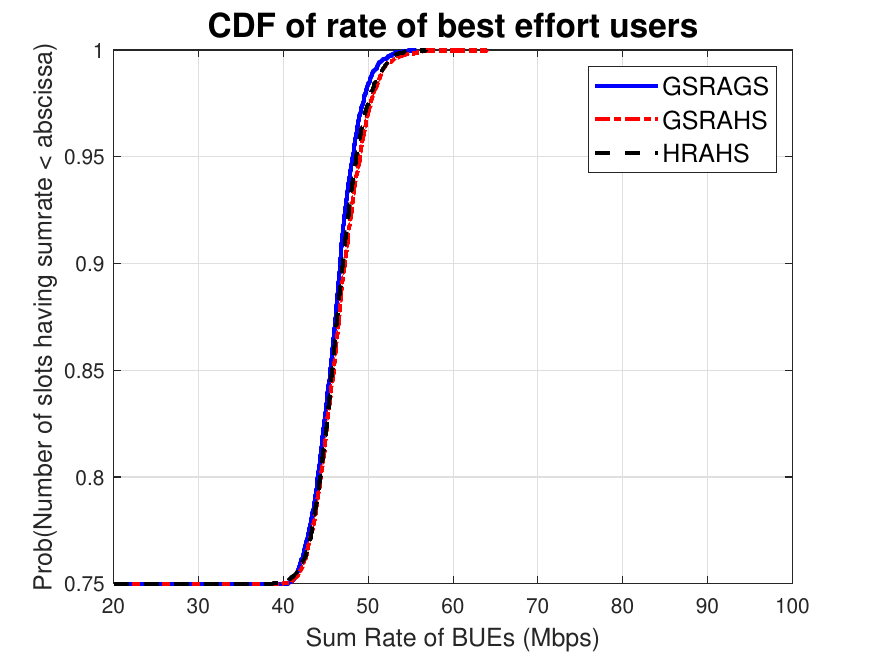}
\caption{Sum Rate of BUEs.}
\label{BUE}
\end{figure}

Figure \ref{BUE} shows the CDF of the sum rate of ten \acp{BUE} across all the numerology zero \acp{RB} of \acp{BWP}-$2$, observed over the entire simulation time window. Since all algorithms use the same resource allocation for \acp{BUE}, hence, the performance of the \acp{BUE} also remains the same. The future work is to design a dynamic \ac{BWP} adaptation mechanism to improve the performance of the \acp{BUE} operating in \ac{BWP}-$2$.

\section{Conclusion}\label{sec:Conclusion}
In this work, we presented a novel resource allocation and sharing algorithm tailored for 5G-based V2X communication systems. Our approach addresses the diverse QoS requirements of V2I and V2V users by leveraging the Gale-Shapley stable matching algorithm for both resource allocation and pairing, resulting in reduced computational complexity. By grouping RBs into RCs and applying a bisection search for power allocation, we efficiently mitigated interference between V2I and V2V users sharing the same radio resources. Simulation results demonstrated that our proposed Gale-Shapley Resource Allocation with Gale-Shapley Sharing (GSRAGS) algorithm achieves at par performance while maintaining low complexity, making it a highly efficient solution for future V2X communication systems.

\bibliographystyle{IEEEtran}
\bibliography{COMSN_2025}

\begin{acronym}
	\acro{2G}{2$^\text{nd}$ Generation}
	\acro{3G}{3$^\text{rd}$ Generation}
	\acro{4G}{4$^\text{th}$ Generation}
	\acro{5G}{5$^\text{th}$ Generation}
	\acro{3GPP}{$\text{3}^\text{rd}$ Generation Partnership Project}
	\acro{A3C}{Actor-Critic}
	\acro{ABR}{Adaptive Bitrate}
	\acro{BS}{Base Station}
	\acro{BUE}{Best-Effort UE}
	\acro{BWP}{Bandwidth Partition}
	\acro{CDN}{Content Distribution Network}
	\acro{CDF}{Cumulative Distribution Function}
	\acro{CSI}{Channel State Information}
	\acro{CUE}{Cellular User Equipment}
	\acro{DASH}{Dynamic Adaptive Streaming over HTTP}
	\acro{DL}{deep learning}
 \acro{DRL}{Deep Reinforcement Learning}
	\acro{DRX}{Discontinuous Reception}
	\acro{D2D}{Device-to-Device}
	\acro{EDGE}{Enhanced Data Rates for \ac{GSM} Evolution.}
			\acro{gNB}{general NodeB}
	\acro{eNB}{evolved NodeB}
	\acro{GSM}{Global System for Mobile}
	\acro{FL}{Federated Learning}
	\acro{TL}{Transfer Learning}
	\acro{HD}{High Definition}
	\acro{HSPA}{High Speed Packet Access}
	\acro{LSTM}{Long Short Term Memory}
	\acro{LTE}{Long Term Evolution}
	\acro{ML}{machine Learning}
	\acro{MTL}{Multi-Task Learning}
	\acro{MCS}{Modulation and Coding Scheme}
	\acro{NSA}{Non-Standalone}
 \acro{NR-V2X}{New Radio-Vehicle-to-Everything}
	\acro{HVPM}{High voltage Power Monitor}
 \acro{PUE}{Pedestrian UE}
 \acro{PDCCH}{Physical Downlink Control Channel}
	\acro{QoS}{Quality of Service}
	\acro{QoE}{Quality of Experience}
	\acro{RB}{Resource Block}
	\acro{RF}{Random Forest}
	\acro{RFL}{Random Forest}
	\acro{RL}{Reinforcement Learning}
	\acro{RRC}{Radio Resource Control}
	\acro{RSSI}{Received Signal Strength Indicator}
	\acro{RSRP}{Reference Signal Received Power}
	\acro{RSRQ}{Reference Signal Received Quality}
	\acro{SCS}{Sub-carrier \ Spacing}
	\acro{SINR}{Signal-to-Interference-Plus-Noise-Ratio}
	\acro{SNR}{Signal-to-Noise-Ratio}
	\acro{SE}{spectral efficiency}
	\acro{UE}{User Equipment}
	\acro{UHD}{Ultra HD}
	\acro{VUE}{Vehicular UE}
	\acro{VoLTE}{Voice over LTE}
	\acro{RNN}{Recurrent Neural Network}
	\acro{WiFi}{Wireless Fidelity}
	\acro{ARIMA}{Auto Regressive Integrated Moving Average}
 \acro{ms}{milliseconds}
	\acro{ML}{Machine Learning}
	\acro{NR}{New Radio}
	\acro{Non-IID} {Independent and Identically Distributed}
 \acro{OFDMA}{Orthogonal Frequency Division Multiple Access}
 \acro{SL}{Sidelink}
	\acro{TP}{Throughput Prediction}
	\acro{CQI}{Channel Quality Indicator}
	\acro{URLLC}{Ultra Reliable Low Latency Communications}
	\acro{V2X}{Vehicle-to-Everything}
	\acro{C-V2X}{Cellular Vehicle-to-Everything}
	\acro{V2I}{Vehicle-to-Infrastructure}
	\acro{V2V}{Vehicle-to-Vehicle}
 \acro{V2P}{Vehicle-to-Pedestrian}
    \acro{OFDM}{Orthogonal Frequency Division Multiplexing }
    \acro{BWP}{Bandwidth Part}
    \acro{PUCCH}{Physical Uplink Control Channel}
    \acro{TTL}{Time-to-Live}
    \acro{RA}{resource allocation}
    \acro{RSU}{Roadside Unit}
    \acro{BLER}{Block Error Rate}
    \acro{LA}{Link Adaptation}
    \acro{RC}{Resource Chunk}
    \acro{TTI}{Transmission Time Interval}
    \acro{HRAHS}{Hungarian Resource Allocation Hungarian Sharing}
    \acro{PLR}{Packet Loss Ratio}
\end{acronym}

\end{document}